\documentclass{webofc}

\usepackage[varg]{txfonts} 
\usepackage{hyperref}
\usepackage{url}
\hypersetup{colorlinks=true,citecolor=blue,urlcolor=blue,linkcolor=blue}

\begin{document}

\title{
 Quantum Anomalies and Parity-odd CFT Correlators for Chiral States of Matter  \footnote{Presented by S. Lionetti at QCD@Work, 18 - 21 June 2024, Trani, Italy\\} }

\author{\firstname{Claudio} \lastname{Corianò}\inst{1,2,3}\fnsep\thanks{\email{claudio.coriano@le.infn.it }} \and
        \firstname{Stefano} \lastname{Lionetti}\inst{1,2}\fnsep\thanks{\email{stefano.lionetti@unisalento.it}}
}

\institute{Dipartimento di Matematica e Fisica, Universit\`{a} del Salento 
	and INFN Sezione di Lecce, Via Arnesano 73100 Lecce, Italy
\and
          National Center for HPC, Big Data and Quantum Computing, Via Magnanelli 2, 40033 Casalecchio di Reno, Italy
\and
            Institute of Nanotechnology,  National Research Council (CNR-NANOTEC), Lecce 73100
          }

\abstract{Chiral currents influence the parity-odd sector of CFT correlators in momentum space, playing a crucial role in the evolution of the quark-gluon plasma in the early universe. We demonstrate that these parity-odd interactions, which couple quarks and gluons to gravitons, can be fully determined in terms of their anomaly content by solving the conformal constraints in momentum space. This process involves a single nonlocal, massless axion-like interaction in the longitudinal channel, which remains protected against thermal and finite density effects.

}
\maketitle

\section{Introduction}
\label{intro}

Interest in studying anomalies in chiral matter, particularly the Adler-Bell-Jackiw (ABJ) anomaly, has grown significantly over the past two decades. This growth spans both condensed matter theory and high-energy physics, focusing on the study of matter under high densities and its behavior in transport. Experimental work, such as heavy ion collision studies, has observed and confirmed anomalous behavior in matter when exposed to strong fields and finite density chiral backgrounds.
The chiral interaction indicates that for massless fermions interacting with electric and magnetic fields \((\vec{E}, \vec{B})\), the chiral fermion number \(N_5\) is not conserved. Instead, it is altered due to the anomaly, expressed as
\begin{equation}
\frac{d N_5}{dt} = \frac{e^2}{2\pi^2} \int d^3x\, \mathbf{E} \cdot \mathbf{B}.
\end{equation}
Axial-vector currents associated with global anomalous \( U(1) \) symmetries can couple to gluons through quark loops, and to gravitons via both quark and gluon loops, facilitated by chiral anomaly interactions. These interactions are crucial in shaping the dynamics of the quark-gluon plasma and become particularly significant in chirally asymmetric environments, where they can intensify chiral asymmetries, especially in the early universe.\\
The phenomenon can be integrated into the magnetohydrodynamic (MHD) equations governing the early universe plasma, leading to the generation of a chiral magnetic cascade. A prominent example of a related phenomenon is the chiral magnetic effect (CME) in the quark-gluon plasma. The CME is characterized by the generation of an electric charge separation along an external magnetic field, driven by an imbalance in chirality caused by the anomaly. In this context, the chiral current is introduced as

\begin{equation}
\vec{J}_5 = \frac{e^2\mu_5 }{2\pi^2}\vec{B},
\end{equation}
where \( \mu_5 = \mu_L - \mu_R \) represents the chiral chemical potential, defined by the difference between the chemical potentials of left-handed (\(L\)) and right-handed (\(R\)) Weyl fermions in the initial state, and \( \vec{B} \) is the magnetic field. The current \( \vec{J}_5 \) acts as an external source that triggers an anomalous response at the quantum level, leading to the generation of an electric field \( \vec{E} \), which is aligned with the magnetic field.\\
This macroscopic phenomenon induces a collective motion within the Dirac sea. The topological nature of the imbalance (\(\mu_5 \neq 0\)) imparts a distinctive quality to the CME current, ensuring its non-dissipative behavior even in the presence of radiative corrections. The presence of a topological protection of the anomaly has been explictly shown in \cite{Coriano:2024nhv,Coriano:2024rcd}.\\
\subsection{Connection with axion physics}
The analysis of parity-odd interactions in the conformal setting may involve pseudoscalar operators, naturally coupled to pseudoscalar fields in a conformal phase of the early universe. \\
Since their introduction as a possible solution of the strong CP-problem in QCD within the Peccei-Quinn theory, 
pseudoscalar fields, particularly those coupled to gravity, can leave distinctive signatures in the stochastic gravitational wave background. Their interaction can be investigated in conformal field theory in momentum space (CFT$_p$).
For axion-like fields, their coupling to gauge fields can induce spin-1 helicity asymmetries before any spontaneous symmetry breaking occurs. A correlator like \( \langle TTO \rangle \), where \( O \) is a pseudoscalar coupled to an axion-like field \( \phi \), is directly linked to a gravitational anomaly. The local effective action in the infrared, when a conformal symmetry-breaking scale \( \Lambda \) is present, takes the form

\[
\mathcal{L}_{axion} \supset \frac{\phi}{\Lambda} R_{\mu\nu} \tilde{R}^{\mu\nu},
\]
where \( R_{\mu\nu} \) is the Ricci tensor. This correlator describes the interaction between the pseudoscalar and two gravitational waves at a semiclassical level in an effective vertex in the forms of local realization. This interaction is analogous to the familiar \( (\phi/f) F \tilde{F} \) coupling in axion physics, which is associated with the \( \langle JJO \rangle \) correlator.

\section{Anomalies and CFT$_p$: Non Lagrangian formulations}
Of all the anomalies, the chiral and conformal ones are among the most discussed. The violation of the chiral symmetry at the quantum level manifests as a breakdown in the conservation of the axial current $J^\mu_5=\bar{\psi}\gamma^\mu\gamma_5 \psi$
\begin{equation}\label{eq:chiranom}
	\nabla_\mu J^\mu_5=a_1\, \varepsilon^{\mu\nu\rho\sigma} F_{\mu\nu}F_{\rho\sigma} + a_2\, \varepsilon^{\mu\nu\rho\sigma} R^\alpha_{\, \, \beta \mu\nu}R^\beta_{\, \,\alpha \rho\sigma}
\end{equation}
where $a_1$ and $a_2$ are numerical coefficients. Initially discovered in the context of high-energy physics, the chiral anomaly's significance extends far beyond. Its implications span from cosmology \cite{Kamada:2022nyt} to condensed matter physics \cite{Chernodub:2021nff}, with phenomena such as the quantum Hall effect, the chiral magnetic effect and applications to topological insulators.\\
On the other hand, the violation of conformal symmetry at the quantum level is reflected by the energy-momentum tensor's failure to be traceless
\begin{equation}\label{eq:confanom}
	g_{\mu \nu}\left\langle T^{\mu \nu}\right\rangle=b_1 E_4+b_2 C^{\mu \nu \rho \sigma} C_{\mu \nu \rho \sigma}+b_3 \nabla^2 R+b_4 F^{\mu \nu} F_{\mu \nu}+f_1 \varepsilon^{\mu \nu \rho \sigma} R_{\alpha \beta \mu \nu} R_{\rho \sigma}^{\alpha \beta}+f_2 \varepsilon^{\mu \nu \rho \sigma} F_{\mu \nu} F_{\rho \sigma} 
\end{equation}
where $ C^{\mu \nu \rho \sigma} $ is the Weyl tensor, $ E_4$ is the Gauss-Bonnet term and $b_i$ and $f_i$ are numerical coefficients. 
The last two terms on the right-hand side of the equation are parity-odd and have been the center of recent discussions (see for example \cite{Bonora:2022izj,Abdallah:2023cdw, Bonora:2021mir,Larue:2023qxw,Bastianelli:2022hmu}). 
The anomalous laws described above translate into Ward identities that dictate the behaviour of various three-point correlation functions.
We are going to present a short overview on parity-odd correlation functions which are affected by such anomalies in the conformal limit. \\
The idea of using conformal Ward identities to determine the structure of three-point functions in momentum space was presented for the first time independently in \cite{Coriano:2013jba} and 
\cite{Bzowski:2013sza}, the second of which outlines a method that includes the tensor case.
Investigations into parity-odd CFT correlators have only recently emerged \cite{Jain:2021gwa,Buchbinder:2023coi,Coriano:2023hts,Coriano:2023cvf,Coriano:2023gxa,Coriano:2024ssu}.\\
CFT$_p$, especially with broken parity, finds applications in cosmology, condensed matter physics, holography and the conformal bootstrap program. Indeed, the recent discovery of applications to topological insulators and Weyl semimetals has sparked new interest in the topic \cite{Coriano:2024ive}.\\

\section{Decomposition of the correlators}
The chiral and conformal Ward identities describe the behavior of the longitudinal and trace part of correlators, respectively. It is therefore useful to decompose a current $J^\mu$ and the energy-momentum tensor $T^{\mu \nu}$ in terms of their transverse-traceless part and longitudinal-trace one (also called "local")
\begin{equation}
\begin{aligned}
	&J^{\mu_i}(p_i)= j^{\mu_i}(p_i)+j_{  loc}^{\mu_i}(p_i),\qquad\qquad T^{\mu_i\nu_i}(p_i)= t^{\mu_i\nu_i}(p_i)+t_{loc}^{\mu_i\nu_i}(p_i)
\end{aligned}
\end{equation}
where
\begin{equation}
\begin{aligned}
	\label{loct}
	&j^{\mu_i}(p_i)=\pi^{\mu_i}_{\alpha_i}(p_i)\,J^{\alpha_i }(p_i), \hspace{1ex}&&j_{ loc}^{\mu_i}(p_i)=\frac{p_i^{\mu_i}\,p_{i\,\alpha_i}}{p_i^2}\,J^{\alpha_i}(p_i),\\
	&t^{\mu_i\nu_i}(p_i)=\Pi^{\mu_i\nu_i}_{\alpha_i\beta_i}(p_i)\,T^{\alpha_i \beta_i}(p_i),\qquad &&t_{loc}^{\mu_i\nu_i}(p_i)=\Sigma^{\mu_i\nu_i}_{\alpha_i\beta_i}(p)\,T^{\alpha_i \beta_i}(p_i),
\end{aligned}
\end{equation}
having introduced the transverse $(\pi)$, transverse-traceless ($\Pi$) and longitudinal-trace ($\Sigma$) projectors, given respectively by 
\begin{equation}
\begin{aligned}
	&\pi^{\mu}_{\alpha}  = \delta^{\mu}_{\alpha} - \frac{p^{\mu} p_{\alpha}}{p^2}, \qquad\qquad
	\Pi^{\mu \nu}_{\alpha \beta}  = \frac{1}{2} \left( \pi^{\mu}_{\alpha} \pi^{\nu}_{\beta} + \pi^{\mu}_{\beta} \pi^{\nu}_{\alpha} \right) - \frac{1}{d - 1} \pi^{\mu \nu}\pi_{\alpha \beta}, \\&
	\Sigma^{\mu_i\nu_i}_{\alpha_i\beta_i}=\frac{p_{i\,\beta_i}}{p_i^2}\Big[2\delta^{(\nu_i}_{\alpha_i}p_i^{\mu_i)}-\frac{p_{i\alpha_i}}{(d-1)}\left(\delta^{\mu_i\nu_i}+(d-2)\frac{p_i^{\mu_i}p_i^{\nu_i}}{p_i^2}\right)\Big]+\frac{\pi^{\mu_i\nu_i}(p_i)}{(d-1)}\delta_{\alpha_i\beta_i}.
\end{aligned}
\end{equation}
As an example, we now consider the $\langle{JJJ_5}\rangle$ correlator, built with two vector currents $J$ and one axial current $J_5$. The vector currents are conserved, while the axial one is not. Therefore, the conservation Ward identities are given by
\begin{equation}\label{eq:jjjconsward3p}
	\begin{gathered}
		p_{1\mu_1}\,\langle{J^{\mu_1}(p_1)J^{\mu_2}(p_2)J_5^{\mu_3}(p_3)}\rangle=0,\qquad\qquad
		p_{2\mu_2}\,\langle{J^{\mu_1}(p_1)J^{\mu_2}(p_2)J_5^{\mu_3}(p_3)}\rangle=0\\
		p_{3\mu_3}\,\langle{J^{\mu_1}(p_1)J^{\mu_2}(p_2)J_5^{\mu_3}(p_3)}\rangle=-8 \, a_1 \, i \, \varepsilon^{p_1p_2\mu_1\mu_2}
	\end{gathered}
\end{equation}
where the last equation follows directly from the chiral anomaly formula \eqref{eq:chiranom}. Due to the equations above, the longitudinal components of the vector currents vanish and the correlator consists only of a transverse part and an anomalous longitudinal part in $J_5$.
Using the projectors mentioned above, we can express the general form for this correlator as
\begin{equation}
	\begin{aligned}
		\langle J^{\mu_1 }(p_1) J^{\mu_2 } (p_2)J_5^{\mu_3}(p_3)\rangle =&\,\langle j^{\mu_1 }(p_1) j^{\mu_2 } (p_2)j_{5\,loc}^{\mu_3}(p_3)\rangle +\langle j^{\mu_1 }(p_1) j^{\mu_2 } (p_2)j_5^{\mu_3}(p_3)\rangle \\&
		=-8\, i  \, a_1 \, \frac{p_{3 \mu_3}}{p_3^2}\varepsilon^{p_1p_2\mu_1\mu_2}+ \pi^{\mu_1}_{\alpha_1}(p_1)\pi^{\mu_2}_{\alpha_2}(p_2)\pi^{\mu_3}_{\alpha_3}(p_3)\, X^{\alpha_1\alpha_2\alpha_3}
	\end{aligned}
\end{equation}
where the transverse-traceless component is parametrized by an undetermined tensor $X^{\alpha_1\alpha_2\alpha_3}$.\\
In general, by imposing (anomalous) Ward identities such as those given in equations  \eqref{eq:chiranom} and \eqref{eq:confanom}, we can fully determine the longitudinal-trace part of a correlator. Specifically, correlators affected by an anomaly exhibit a pole in the anomalous momentum within the longitudinal-trace part.
On the other hand, the transverse-traceless component of a correlator is not directly constrained by these Ward identities. In order to fix such component, we need to impose the conformal constraints following from the invariance under dilatations and special conformal transformations. 
In the next section, we will briefly outline the procedure.

\section{Conformal Ward identities}
The conformal constraints consist of differential equations that a correlator need to satisfy. By solving such equations, we can determine the most general expression of a correlator in CFT$_p$.\\
Consider an arbitrary $n$-point correlator $\left\langle \ldots\right\rangle$ with the overall delta function for momentum conservation removed.
In momentum space, the invariance of a correlator under dilatations is reflected in the following first-order differential equation
\begin{equation*}
	\left[\sum_{j=1}^n \Delta_j-(n-1) d-\sum_{j=1}^{n-1} p_j^\alpha \frac{\partial}{\partial p_j^\alpha}\right] 
	\left\langle \ldots\right\rangle
	=0
\end{equation*}
while the invariance under special conformal transformations lead to a second-order differential equation 
\begin{equation*}
	\begin{aligned}
		0 & =\mathcal{K}^\kappa
		\left\langle \ldots\right\rangle
	\end{aligned}
\end{equation*}
where
\begin{equation*}\small
	\begin{aligned}
		\mathcal{K}^\kappa & \equiv \sum_{j=1}^{n-1}\left(2\left(\Delta_j-d\right) \frac{\partial}{\partial p_j^\kappa}-2 p_j^\alpha \frac{\partial}{\partial p_j^\alpha} \frac{\partial}{\partial p_j^\kappa}+\left(p_j\right)_\kappa \frac{\partial}{\partial p_j^\alpha} \frac{\partial}{\partial p_{j \alpha}}\right)+ \mathcal{K}^\kappa _{spin}
	\end{aligned}
\end{equation*}
The expression of the operator $\mathcal{K}^\kappa _{spin}$ depends on the tensorial structure of the correlator \cite{Bzowski:2013sza}.
The general solution of these equations can be expressed in terms of integrals involving a product of three Bessel functions (‘3K integrals’)
\begin{equation*}
	I_{\alpha\left\{\beta_1 \beta_2 \beta_3\right\}}\left(p_1, p_2, p_3\right)=\int d x x^\alpha \prod_{j=1}^3 p_j^{\beta_j} K_{\beta_j}\left(p_j x\right)
\end{equation*}
where $K_\nu$ is a modified Bessel function of the second kind 
\begin{equation*}
	\begin{aligned}
		&K_\nu(x)=\frac{\pi}{2} \frac{I_{-\nu}(x)-I_\nu(x)}{\sin (\nu \pi)}, \qquad\qquad \nu \notin \mathbb{Z} \\& I_\nu(x)=\left(\frac{x}{2}\right)^\nu \sum_{k=0}^{\infty} \frac{1}{\Gamma(k+1) \Gamma(\nu+1+k)}\left(\frac{x}{2}\right)^{2 k}
	\end{aligned}
\end{equation*}
Details can be found in \cite{Bzowski:2013sza}, including the connection to ordinary Feynman integrals for free field theory realizations. The analysis of the parity-odd sector in the presence of chiral and conformal anomalies is discussed in \cite{Coriano:2024ssu,Coriano:2023cvf,Coriano:2023hts,Coriano:2023gxa}.

\section{Solution of the CWIs: the correlators $\langle JJJ_5\rangle$, $\langle TTJ_5\rangle$ and $\langle J_5J_5J_5 \rangle$}
\begin{figure} 
	\centering
	\includegraphics[scale=0.4]{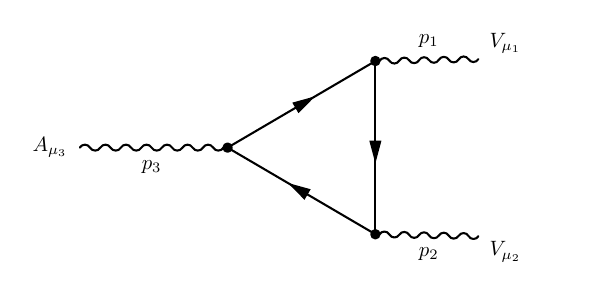}
	\caption{Feynman diagram for the chiral anomaly interaction $\langle JJJ_5\rangle$}
	\label{Fig:vva}
\end{figure}
\begin{figure}
	\centering
	\includegraphics[scale=0.34]{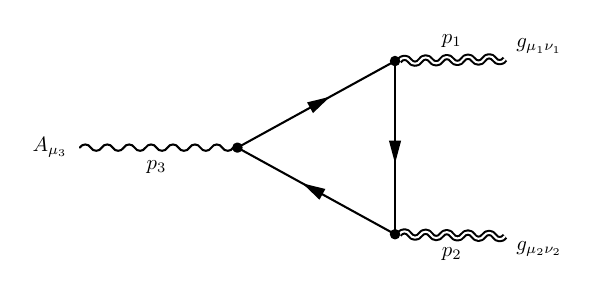}
	\includegraphics[scale=0.34]{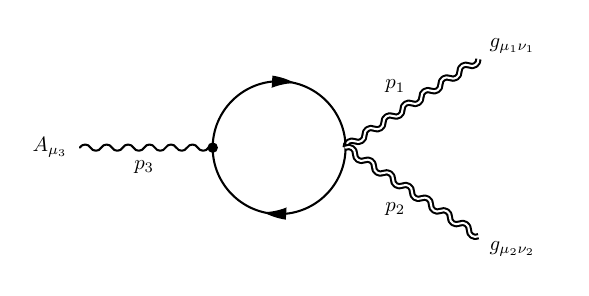}
	\includegraphics[scale=0.34]{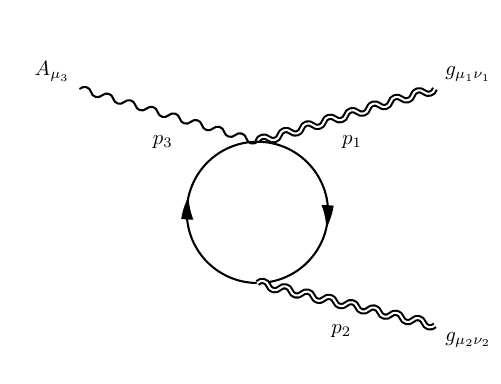}
	\includegraphics[scale=0.34]{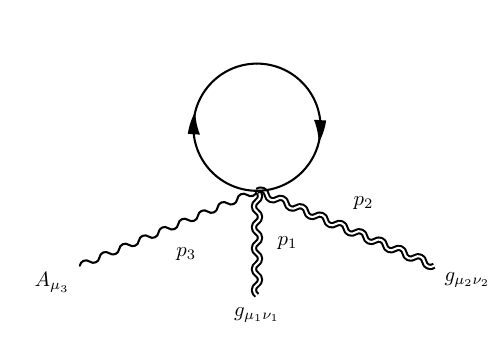}
	\caption{Topologies of Feynman diagrams contributing to the gravitational anomaly interaction $\langle TTJ_5\rangle$. }
	\label{Fig:tta}
\end{figure}
We now focus on the correlators that are affected by the chiral anomaly given in eq$.$ \eqref{eq:chiranom}. The first term on the righ-hand side of this equation can be analyzed by looking at the $\langle JJJ_5\rangle$. 
By imposing the conformal Ward identities described in the previous section, we can completely fix the correlator in the following form
\begin{equation}\label{eq:jjj5sol}
	\begin{aligned}
		\langle J^{\mu_1}&\left(p_1\right) 
		J^{\mu_2}\left(p_2\right) J^{\mu_3}_5 	\left(p_3\right)\rangle =-8\, i  \, a_1 \, \frac{p_{3 \mu_3}}{p_3^2}\varepsilon^{p_1p_2\mu_1\mu_2}+ \\& \pi^{\mu_1}_{\alpha_1}\left(p_1\right)
		\pi^{\mu_2}_{\alpha_2} \left(p_2\right) \pi^{\mu_3}_{\alpha_3}
		\left(p_3\right) \, 8 \, i \, a_1 \,  \Bigl[  
		I_{3\{1,0,1\}}\, p_2^2 \, \varepsilon^{p_1 \alpha_1\alpha_2\alpha_3}  -
		I_{3\{0,1,1\}}\, p_1^2 \, \varepsilon^{p_2\alpha_1\alpha_2\alpha_3}  
		\Bigr]
	\end{aligned}
\end{equation}
On the other hand, if we want examine the effect of the second term in eq. \eqref{eq:chiranom} we need to consider the $\langle TTJ_5\rangle$ correlator. By solving the conformal Ward identities, we can determine its general expression
\small
\begin{equation}\label{eq:ttj5sol}
	\begin{aligned}
		&\left\langle T^{\mu_{1} \nu_{1}} T^{\mu_{2} \nu_2} J_5^{\mu_{3}}\right\rangle=4 i a_2 \frac{p_3^{\mu_3}}{p_3^2} \, (p_1 \cdot p_2) \left\{ \left[\varepsilon^{\nu_1 \nu_2 p_1 p_2}\left(g^{\mu_1 \mu_2}- \frac{p_1^{\mu_2} p_2^{\mu_1}}{p_1 \cdot p_2}\right) +\left( \mu_1 \leftrightarrow \nu_1 \right) \right] +\left( \mu_2 \leftrightarrow \nu_2 \right) \right\}+\\&\hspace{0.5cm}
		\Pi_{\alpha_{1} \beta_{1}}^{\mu_{1} \nu_{1}}\left({p}_{1}\right) \Pi_{\alpha_{2} \beta_{2}}^{\mu_{2} \nu_{2}}\left({p}_{2}\right) \pi_{\alpha_{3}}^{\mu_{3}}\left({p_3}\right) \bigg[
		A_1\varepsilon^{p_1\alpha_1\alpha_2\alpha_3}p_2^{\beta_1}p_3^{\beta_2}
		+A_2\varepsilon^{p_1\alpha_1\alpha_2\alpha_3}\delta^{\beta_1\beta_2}
		+\bigg(\big\{p_1,\alpha_1,\beta_1\big\} \leftrightarrow \big\{p_2, \alpha_2,\beta_2\big\}\bigg)
		\bigg]
	\end{aligned}
\end{equation}
\normalsize
with
\begin{equation}\label{eq:resultffconformi}
	\begin{aligned}
		&A_1=-4\, i \, a_2\, p_2^2\, I_{5\{2,1,1\}}\\
		&A_2=-8\, i \, a_2\, p_2^2\, \bigg(p_3^2\, I_{4\{2,1,0\}}-1 \bigg).
	\end{aligned}
\end{equation}
\\
Let us now comment on the equations \eqref{eq:jjj5sol} and \eqref{eq:ttj5sol}
\begin{itemize}
	\item Both formulas include, in their first row, an anomalous term in the longitudinal sector of the correlator, characterized by an anomalous pole $1/p_3^2$. In their second row, the formulas feature the transverse-traceless component, which is fixed in terms of 3K integrals by imposing conformal invariance.
	\item At first glance, one might assume that the chiral anomaly only affect the longitudal component of correlators, expecially when looking at equations like \eqref{eq:jjjconsward3p}. However, this assumption does not hold true in a CFT, where the transverse component can indirectly depend on the anomaly content. This occurs because
	the conformal equations couple the longitudinal and transverse sectors, thereby leading to the appearance of the anomalous coefficients $a_1$ and $a_2$ in the transverse-traceless component as well. 
	\item The correlators are fully determined by the anomalous coefficient $a_1$ or $a_2$, with no additional degrees of freedom. This is unusual for parity even 3-point correlators in a CFT which typically depend on multiple unfixed constants. Here, the parity-odd correlators are  generated entirely by an anomaly and they vanish if the anomaly is absent ($a_1 = 0$, $a_2 = 0$).
	\item
	An alternative approach for studying this topic is given by the perturbative realization of CFT correlators which allows us to handle more simplified expressions proceeding with analyzing an ordinary Feynman expansion (see Fig$.$ \ref{Fig:vva} and \ref{Fig:tta}).
	The perturbative analysis serves as an important independent check for the non-perturbative conformal methods described in this paper and it also provides information on the emergence of the conformal and chiral anomalies. As demonstrated in \cite{Coriano:2023hts,Coriano:2023gxa}, the equations \eqref{eq:jjj5sol} and \eqref{eq:ttj5sol} are in perfect agreement with the perturbative results.
\end{itemize}
One can repeat the calculations for different correlators. In the case of the $	\left\langle J_5 J_5J_5\right\rangle$, where all the currents are anomalous, solving the conformal Ward identities yields the following result
\begin{equation}
	\left\langle J_5^{\mu_1} J_5^{\mu_2} J_5^{\mu_3}\right\rangle=\frac{1}{3}\bigg[\left\langle J_5^{\mu_1} J^{\mu_2} J^{\mu_3}\right\rangle+\left\langle J^{\mu_1} J_5^{\mu_2} J^{\mu_3}\right\rangle+\left\langle J^{\mu_1} J^{\mu_2} J_5^{\mu_3}\right\rangle\bigg]
\end{equation}
where the explicit form of the terms on the right-hand side can be obtained by appropriately permuting the indices and momenta in equation \eqref{eq:jjj5sol}.\\
The methodology described in this paper can also be applied to correlators affected by trace anomalies. Such phenomena are discussed in \cite{Coriano:2023cvf}.

\subsection{Conclusions and outlook}
In recent years, it has become increasingly clear that chiral anomalies—resulting from the breaking of classical global symmetries by quantum corrections—play a pivotal role in the dynamics of fundamental interactions. This influence extends not only to the vacuum case \((\mu = 0)\), but also to scenarios involving chiral chemical potentials, as demonstrated in the context of the chiral magnetic effect. The potential for experimental verification of these interactions, which could reveal the nature of the pseudoparticles emerging from virtual corrections, as predicted by recent and past analyses, should be taken seriously at the experimental level. We have briefly outlined how conformal symmetry and a nonlocal interaction in a longitudinal channel determine the interactions underlying chiral and gravitational anomalies in gauge theories.
\vspace{0.5cm}\\
\centerline{\bf Acknowledgements}
This work is partially funded by the European Union, Next Generation EU, PNRR project "National Centre for HPC, Big Data and Quantum Computing", project code CN00000013; by INFN, inziativa specifica {\em QG-sky} and by the grant PRIN 2022BP52A MUR "The Holographic Universe for all Lambdas" Lecce-Naples.


\end{document}